\documentclass[12pt]{article}
\usepackage{amssymb}
\usepackage{epsfig}
\usepackage{graphicx}
\usepackage{amsmath}
\usepackage{verbatim}
\begin{document}

\textwidth 6.3in \textheight 8.8 in \hoffset -0.3 in \voffset -0.3in
\renewcommand{\baselinestretch}{1.3}
\renewcommand{\thepage}{\arabic{page}}
\renewcommand{\theequation}{\thesection.\arabic{equation}}
\csname @addtoreset\endcsname{equation}{section}
 
\begin{titlepage}
\title{\bf\Large Higgs-Portal Scalar Dark Matter: Scattering Cross Section and Observable Limits \vspace{18pt}}

\author{\normalsize Huayong Han and Sibo Zheng \vspace{12pt}\\
{\it\small   Department of Physics, Chongqing University, Chongqing 401331, P.R. China}\\}

\date{}
\maketitle \voffset -.3in \vskip 1.cm \centerline{\bf Abstract}
\vskip .3cm
The simplest Higgs-portal dark matter model is studied in the light of dark matter self-interacting effects 
on the formation of large scale structures. 
We show the direct detection limits in both the resonant and large mass region.
Finally, we also compare these limits with those at the LHC and Xenon 1T experiments.

 \thispagestyle{empty}

\end{titlepage}
\newpage

\section{Introduction}
The observed dark matter (DM) from galaxy rotation curves requires extension beyond the Standard Model (SM), 
in which there is no viable candidate.
Among other things,
one of the simplest DM models corresponds to coupling the DM sector to SM sector, 
with the SM Higgs scalar as the interaction mediator.
This scenario is known as the Higgs-portal DM.
The direct detection limits at the LUX experiment have excluded a fermion-like but still allow a scalar-like DM 
within mass range between 1 GeV and 10 TeV.

In the minimal version of  Higgs-portal scalar DM \cite{Zee,0702143,0003350,0011335} 
there are only three model parameters, 
which include the DM mass $m_s$, 
the Yukawa coupling constant $\kappa_s$ between DM and the SM Higgs,
and the DM self-interaction coupling constant $\lambda_s$.
The signals of direct or indirect detection in this model are very predictive.
\begin{itemize}
\item $\mathbf{Indirect~detection}$ mainly includes limits on DM annihilation 
into $e^{+}e^{-}$ at PAMELA \cite{0810.0713, 0911.3898,1005.4678}, 
into $\gamma$ rays at Fermi-LAT \cite{1501.03507, 1508.04418,1510.07562}, 
neutrinos in the sun \cite{1505.03781,WYL,1312.0797}, 
and Higgs invisible decay for the DM mass below half of the Higgs mass $m_{h}$.

\item $\mathbf{Direct~detection}$ mainly includes limits on the DM-nucleon spin-independent scattering at Xenon 100 \cite{Xenon100} and LUX \cite{LUX, 1507.04007},  and the direct production at hadron  \cite{1112.3299,1412.0258, 1407.6882, 1601.06232} and lepton \cite{1311.3306} colliders.
\end{itemize}
Summarizing experimental limits above, 
the scalar DM mass is tightly constrained to two regions 
\footnote{If the Hubble parameter $H$ during inflation is above $10^{16}$ GeV,
the resonant mass region is totally excluded \cite{1509.01765}.
In contrast, these two regions are both consistent with present experimental limits if $H$ is small enough.
In this letter, we take the later assumption.}, 
\begin{eqnarray}{\label{regions}}
\text{resonant mass region} &:& 62.5 ~\text{GeV}  \leq m_{s}\leq  66 ~\text{GeV}, \nonumber\\
\text{large mass region} &:&  m_{s}\geq 185~\text{GeV}.
\end{eqnarray}

In this paper we explore direct detection on this model 
via DM self-interacting effects on the formation of large scale structures (LSS) \cite{1310.7945}, 
which is less studied in comparison with the DM-nucleon spin-independent scattering.
As firstly described by Spergel and Steinhardt \cite{9909386},
self-interacting DM may be used to explain the constant core problem \cite{0002409,0809.0898,1201.5892,1208.3025, 1412.1477} 
and missing satellites in DM halos at the dwarf scale 
\footnote{The number of DM halos at this scale is  roughly about $\sim 1000$ as inferred either from simulation \cite{simulation} 
or analytic theory \cite{analytic}, but less than $\sim 100$ galaxies are observed \cite{Observation}.}.
In the former one, kinetic energy is transmitted from the hot outer halo inward because of DM self-interaction,
with suitable strength (as described by $\sigma/m_s$, here $\sigma$ denotes the scattering cross section for $ss\rightarrow ss$ ).
In the later case, 
DM self-interaction could lead to satellite evaporation due to the DM particles within the satellites 
being kicked out by high-velocity encounters with DM particles from the surrounding dark halo of the parent galaxy. 

Table \ref{limit} shows present limits on $\sigma/m_s$ based on astrophysical observations at different galactic scales.
The studies of DM self-interacting effects on the formation of LSS will shed light on two aspects.
At first, the DM self-interaction coupling constant $\lambda_s$ is constrained more efficiently,
in compared with constraints arising from the DM relic density, direct detection limits at the LUX or LHC,
which have little relevance to $\lambda_s$.
Also, it provides the limits for discovery of DM in terms of astrophysical observations.

\begin{table}
\begin{center} 
\begin{tabular}{|c|c|c|c|}
  \hline
   Galactic Scale & limit ($\text{cm}^{2}/\text{g}$) & velocity (km/s) & Refs. \\
  \hline\hline
 Milky Way & $\sigma/m_{s}\leq 1.0$ & $\sim 10^{2}$ & \cite{1311.6524} \\
 Cluster & $\sigma/m_{s}\leq 1.25$ & $\sim 10^{3}$ &\cite{1208.3026} \\
  \hline
\end{tabular}
\caption{Upper bounds on $\sigma/m_{s}$ at different galactic scales inferred from DM self-interacting effects on the formation of LSS. In comparison with \cite{1208.3026},
recent observations of cluster collisions \cite{science} give rise to a slightly stronger upper bound $\sigma/m_{s}\leq 0.47$. }
\label{limit}
\end{center}
\end{table}

The rest of this paper is organized as follows.
In Sec. 2, we calculate the tree-level value for DM scattering cross section $\sigma_{0}$ 
in terms of Madgraph5 \cite{code} and Feynman rules generator \cite{FeynRules} .
We eliminate parameter $\kappa_s$ via the constraint from DM relic abundance,
therefore $\sigma_0$ only depends on the remaining parameters $m_s$ and $\lambda_s$.
In Sec. 3, we consider the Sommerfeld effect \cite{Sommerfeld} on DM scattering cross section 
due to the DM self-interaction \cite{0902.0688, 0903.5307, 0910.5713}.
In comparison with a massless or light-mass mediator, 
the Higgs mass upper bounds the enhancement factor significantly.
The enhancement on the DM scattering cross section is verified to be mild in the resonant mass region, 
and less than $\sim 10^{4}-10^{5}$ in the large mass region for $m_s$ above $\sim 2$ TeV.
In Sec. 4 we compare the experimental limits with those at the LHC and Xenon experiments.
Finally, we conclude in Sec. 5.

\begin{figure}
\centering
\includegraphics[width=4.5in]{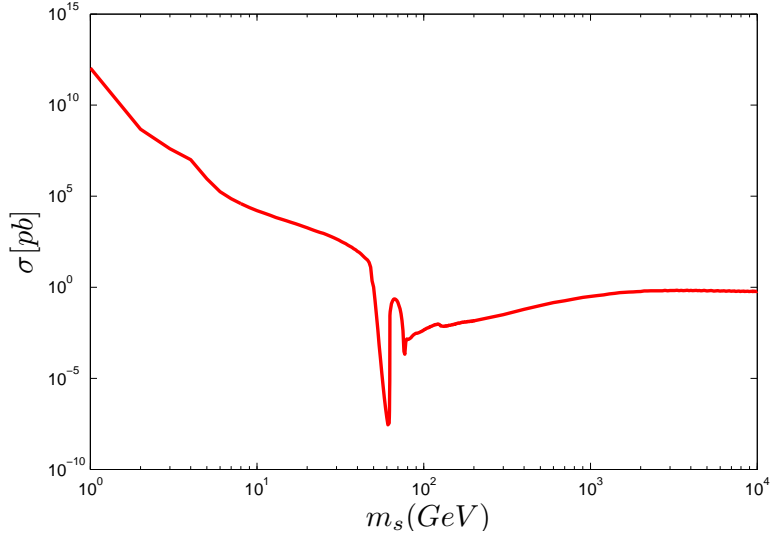}
\caption{The dependence of tree-level scattering cross section $\sigma_0$ on DM mass without DM self-interaction ($\lambda_{s}=0$). 
Note that the dependence on $\kappa_s$ is eliminated via the measured DM relic abundance \cite{1303.5076}, 
which is numerically calculated via MicrOMEGAs \cite{1407.6129}.}
\label{s1}
\end{figure}

\section{Tree-Level Scattering cross section}
The relevant Lagrangian for our model is given by,
\begin{eqnarray}{\label{Lagrangian1}}
\mathcal{L}=-\frac{1}{2}\left(\partial s\right)^{2}
+\frac{\mu^{2}_{s}}{2}s^{2}+\frac{\lambda_{s}}{2}s^{4}+\frac{\kappa_{s}}{2}s^{2}\mid H\mid^{2}.
\end{eqnarray}
Expand DM field $s$ and Higgs field $h$ along their vacuum expectation value $\left<s\right>=0$ 
and $\left<H\right>=(\upsilon_{\text{EW}}+h)/\sqrt{2}$, respectively, we obtain
\begin{eqnarray}{\label{Lagrangian2}}
\mathcal{L}= -\frac{1}{2}\left(\partial s\right)^{2} +\frac{1}{2} m_{s}^{2} s^{2} +\frac{\lambda_{s}}{2}s^{4}
+\frac{\kappa_{s}\upsilon_{\text{EW}}}{2}s^{2}h+\frac{\kappa_{s}}{4}s^{2}h^{2},
\end{eqnarray}
where $m^{2}_{s}=\mu_{s}^{2}+\kappa_{s}\upsilon_{\text{EW}}^{2}/2$,
and the electroweak scale $\upsilon_{\text{EW}}=246$ GeV.

\begin{figure}
\centering
\includegraphics[width=4.5in]{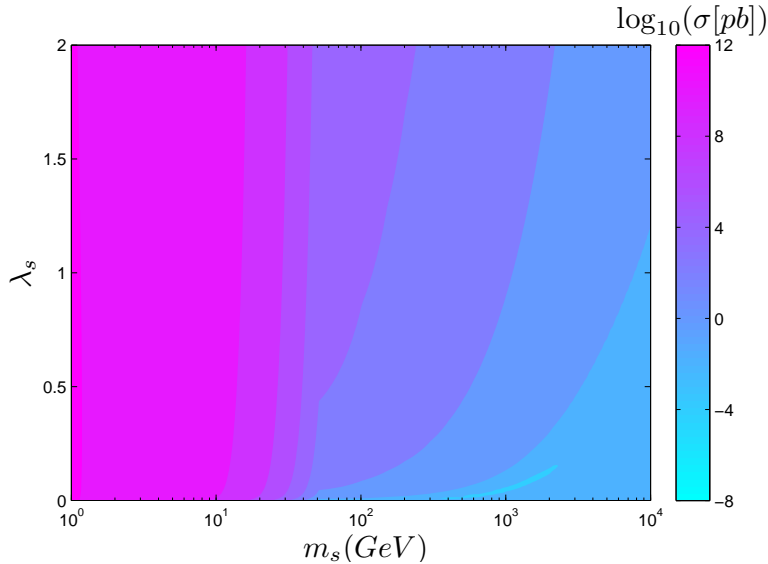}
\caption{Similar to Fig.\ref{s1} but with DM self-interaction included.
Comparison with Fig.\ref{s1}  indicates that large $\lambda_s$ dominates the contribution to $\sigma_{0}$ in the large mass region.}
\label{s2}
\end{figure}

The contributions to DM scattering cross section include two types of Feynman diagrams 
- one with intermediate Higgs scalar field and the other with contact interaction.
The tree-level value for $\sigma_0$ without and with quartic interaction is shown in 
Fig.\ref{s1}  and Fig.\ref{s2}, respectively,  
by using Madgraph5 \cite{code}.
In Fig.\ref{s1} the dependence of $\sigma_0$ on $\kappa_s$ is eliminated in terms of the measured DM relic abundance.
Consequently, the total contribution to $\sigma_0$, as shown in Fig.\ref{s2},
can be presented in the parameter space of $m_s$ and $\lambda_s$.
These numerical values are compatible with analytic approximations in different mass limits \cite{1505.01793},
\begin{equation}{\label{crosssection}}
\frac{\sigma}{10^{4}}[\text{pb}]\sim
 \left\{
\begin{array}{lcl}
1.4\cdot \lambda_{s}^{2}\cdot\left(\frac{100~ \text{GeV}}{ m_{s}}\right)^{2} , 
~~~~~~~~~~~~~~ m_{s}>>m_{h},\\
5.6\cdot \left(\frac{\lambda_{s}}{2}-\frac{\kappa^{2}_{s}\upsilon^{2}}{8m^{2}_{h}}\right)^{2} \cdot\left(\frac{100~ \text{GeV}}{ m_{s}}\right)^{2},   ~m_{s}<<m_{h}.
\end{array} \right. 
\end{equation}

Fig.\ref{s2} indicates that $\sigma_{0}$ is upper bounded as 
$\sigma\leq 10^{12}$ pb in the whole range $0\leq \lambda_{s}\leq 2$, 
which implies \footnote{There is a useful relation among different units: 
1 $\text{cm}^{2}/g$=$1.8\times 10^{12} \text{pb/GeV}=4.62 \times 10^{3}\text{GeV}^{-3}$.}
that $\sigma/m_{s}\leq 1~\text{cm}^{2}/\text{g}$ for $m_s$ above $1$ GeV. 
Although small $\sigma$ is compatible with the limits shown in Table \ref{limit},
relative larger $\sigma$ is more favored in the light of direct detection at further astrophysical observations.
As we will show in the next section, 
the Sommerfeld effect enhaces the magnitude of $\sigma_0$,
which is as large as of order $\sim 10^{4}-10^{5}$ in the large mass region.
It seems that the discovery potential for large mass region can be improved.
This issue will be discussed in detail in Sec. 4.\\

\section{Sommerfeld Effects}
The $S$-wave annihilation cross section for two DM particles moving at small relative velocities, 
is enhanced by a factor ($S$) depending on the inverse velocity $v\sim10^{-3}$,
in compared with $v\sim 0.3$ at the freeze-out time.
This enhancement is known as the Sommerfeld effect,
which corresponds to the summation of a series of ladder diagrams with the mediator repeatedly exchanged.
Since firstly applied to the wino dark matter \cite{0610249}, 
it has been clear that the DM annihilation cross section may be significantly differs from the
DM scattering cross section when these two cross sections are both $S$-wave dominated.

\begin{figure}
\centering
\includegraphics[width=4.5in]{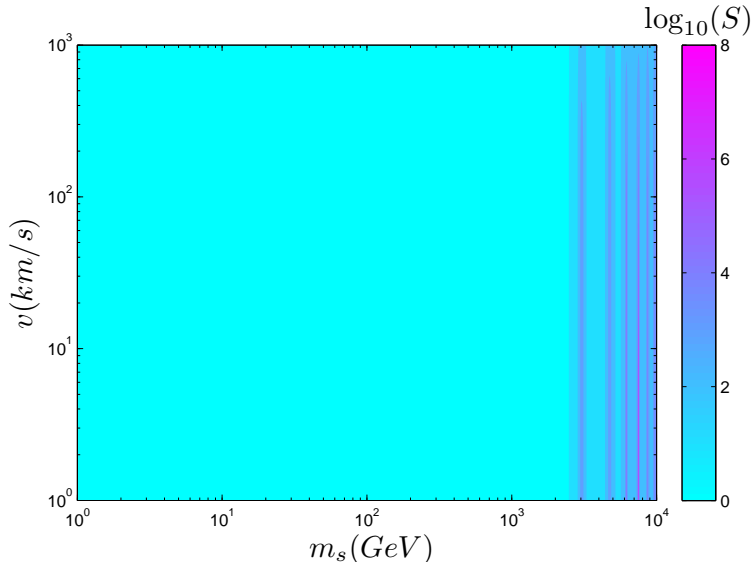}
\caption{Sommerfeld enhancement factor in the parameter space of $m_s$ and $v$.
Note that the dependence of S factor on parameter $\kappa_{s}$ is eliminated by the requirement of DM relic abundance similar to previous treatments. }
\label{s}
\end{figure}

This mediator is the Higgs scalar in our model.
By following the works in \cite{0902.0688,0903.5307,0910.5713}, 
one obtains  the enhancement factor in terms of solving the non-relativistic schrodinger equation,
\begin{eqnarray}{\label{yukawa}}
-\frac{1}{m_{s}}\frac{d^{2}\chi}{dr^{2}}+V(r)\chi =m_{s}v^{2}\chi
\end{eqnarray}
where in our case the Yukawa potential \footnote{Ref. \cite{0801.3440} has considered a similar model.
The Yukawa potential therein differs from ours due to different conventions.},
\begin{eqnarray}{\label{yukawa}}
V(r)=-\frac{\kappa^{2}_{s}}{4\pi r} \exp{(-m_{h}r)}
\end{eqnarray}
The boundary conditions are given by 
\begin{eqnarray}
\chi'(r)&=&im_{s}v\chi(r),\nonumber\\
\chi(r)\mid_{r\rightarrow \infty} &\rightarrow& \exp{(im_{s}vr)}.  
\end{eqnarray}
Under such notation, the Sommerfeld enhancement factor $S$ reads as,
\begin{eqnarray}{\label{S}}
S=\frac{\mid\chi(\infty)\mid^{2}}{\mid\chi(0)\mid^{2}}
\end{eqnarray}

$S$ depends only on parameters $v$ and $m_{s}$, 
as its dependence on $\kappa_{s}$ can be eliminated by the requirement of DM relic abundance.
Fig.\ref{s} shows our numerical solution to the Sommerfeld enhancement factor $S$ in the parameter space of $m_s$ and $v$. 
$S$ is around unity for DM mass below $2$ TeV,
and its maximal value is about $\sim 10^{5}-10^{7}$ for $m_{s} \geq 3$ TeV.
These numerical results agree with the analytic approximation \cite{0903.5307, 0910.5713, 1005.4678},
\begin{eqnarray}{\label{analytic}}
S=\frac{\pi}{\epsilon_{v}}\frac{\sinh\left(\frac{2\epsilon_{v}}{\pi^{2}\epsilon_{s}/6}\right)}{\cosh\left(\frac{2\epsilon_{v}}{\pi\epsilon_{s}/6}\right)-\cos\left(2\pi \sqrt{\frac{1}{\pi^{2}\epsilon_{s}/6}-\frac{\epsilon^{2}_{v}}{(\pi^{2}\epsilon_{s}/6)^{2}}}\right)},\nonumber
\end{eqnarray}
where $\epsilon_{v}=v/\alpha_{\kappa_{s}}$ and $\epsilon_{s}=m_{h}/(\alpha_{\kappa_{s}}m_{s})$.
Although it is not obvious in Fig.\ref{s}, 
we have also verified that $S$ decreases as the velocity $v$ increases.

\begin{figure}
\centering
\includegraphics[width=4.5in]{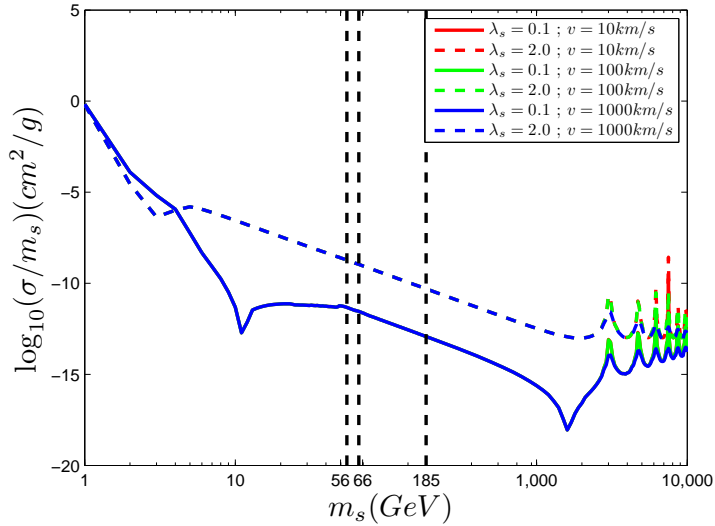}
\caption{$\sigma/m_{s}$ as function of $m_s$ for $\lambda_{s}$ between $0.1$ and $2$ and $v=\{10, 10^{2}, 10^{3}\}$ km/s.
Similar to Fig.\ref{s} the dependence on parameter $\kappa_{s}$ is eliminated.
Note that both the red and green lines overlap with the blue ones for DM mass below $\sim 3$ TeV. }
\label{limits}
\end{figure}

\section{Comparison with LHC and Xenon 1T}
Combining the Sommerfeld effect in the previous section gives rise to our final result on the DM scattering cross section,
\begin{eqnarray}{\label{sigma}}
\sigma= S(\upsilon, m_{s}) \sigma_{0}(\lambda, m_{s}).
\end{eqnarray}
In terms of Fig.\ref{s} we plot $\sigma/m_{s}$ as function of $m_s$ for different $\lambda_s$ and velocity $v$ in the range of $10-10^{3}$ km/s  in Fig.\ref{limits}.
It is shown that the resonant mass region for large $\lambda_{s}\sim 2$ 
can be probed for $\sigma/m_{s}$ of order $\sim 10^{-7}$ $\text{cm}^{2}/g$;
and $\sigma/m_{s}$ of order $\sim 10^{-11}$ $\text{cm}^{2}/g$ is required for small $\lambda_{s} \sim 0.1$.
Smaller limits on $\sigma/m_{s}$ are required for the detection for either smaller $\lambda_s$ or larger DM mass.

Obviously, the value of $\sigma/m_{s}$ is consistent with present astrophysical limits shown in Table \ref{limit} in the whole mass region.
It is also obvious that the simplest Higgs-portal DM model can not provide $\sigma$ large enough to 
explain the puzzles at the dwarf scale as mentioned in the introduction.

The required limits on $\sigma/m_{s}$ for detection seems too small in compared with present limits (of order $\sim 10^{-1}$ $\text{cm}^{2}/g$ ). 
Does it imply that the astrophysical observations on LSS  are less efficient in compared with other direct detection facilities ?
Let us compare these limits with those at the future LHC and Xenon 1T experiments as required for discovery. 
The main observations are summarized in Table \ref{summary}.
See what follows for explanation. \\

\begin{table}
\begin{center} 
\begin{tabular}{|c|c|c|c|}
  \hline
  DM Mass (GeV) & LHC  & Xenon 1T &  LSS\\
  \hline\hline
  \text{Resonant mass region} & $\surd$ & $\times$& $\surd$ \\
 \text{Large mass region} ($185 \leq m_{s}< 3000$) & $\times$ & $\surd$ & $\times$ \\
 \text{Large mass region} ($m_{s}\geq 3000$) & $\times$ & $\times$ & $\times$\\
  \hline
\end{tabular}
\caption{Prospect for the discovery of DM at different experimental facilities. 
We have assumed that the limit on $\sigma/m_{s}$ of order $\sim 10^{-7}$ $\text{cm}^{2}/g$  can be reached in the further astrophysical observations on LSS.
In comparison with the required integrated luminosity $\mathcal{L}$ at least of order $10^{3}$ $\text{fb}^{-1}$ at the 14-TeV LHC,
LSS provides a complementary way to detect the resonant mass region. }
\label{summary}
\end{center}
\end{table}

\begin{figure}
\centering
\includegraphics[width=4.5in]{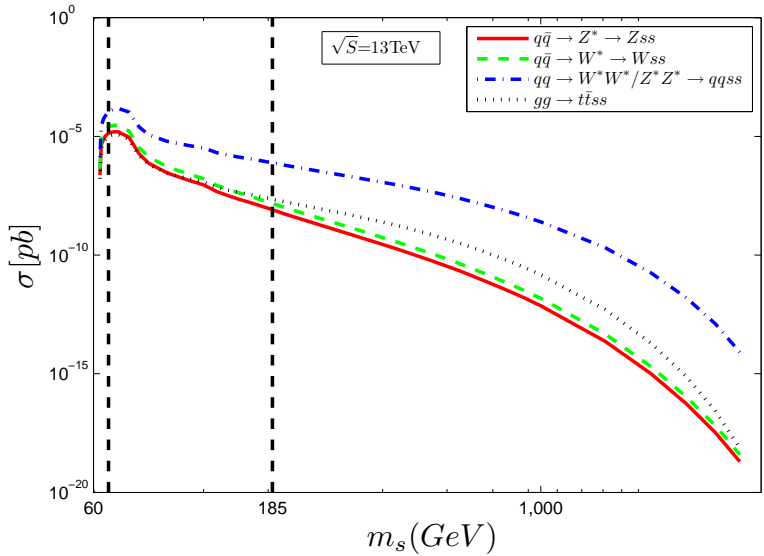}
\caption{Contributions to the production cross section for DM at the 13-TeV LHC,
which are dominated by the vector boson fusion processes.
The dependence on $\kappa_s$ is eliminated similar to Fig. \ref{s1}. }
\label{lhc}
\end{figure}

$(i)$ As shown in Fig.\ref{lhc}, the production cross section for DM at the 13-TeV LHC 
is less than $\sim 10^{-1}$ fb and $10^{-4}$ fb in the resonant mass region and the large mass region respectively.
Therefore the later case is beyond the reach of LHC, and the former case can be detected only for extremely large integrated luminosity $\mathcal{L}$ at least of order $\sim 10^3$ $\text{fb}^{-1}$ if one takes care of the SM background \cite{1601.06232}.\\

$(ii)$ The DM-nucleon spin-independent scattering cross section are less than 
$\sim 10^{-13}$ pb and  $\sim 10^{-8}$ pb in the resonant mass region and large mass region with $m_{s}\geq 3$ TeV, respectively,
which are both beyond the reach of Xenon 1T experiment \cite{1509.01765}.\\

$(iii)$ With the assumption that the limit on $\sigma/m_{s}$ of order $\sim 10^{-7}$ $\text{cm}^{2}/g$ can be reached in the further astrophysical observations on LSS,
the resonant mass region with large $\lambda\sim 2$ can be totally detected.
Therefore, in comparison with the required integrated luminosity $\mathcal{L}$ of order $10^{3}$ $\text{fb}^{-1}$ at the 14-TeV LHC,
LSS provides a complementary way to detect the resonant mass region.

\section{Conclusions}
In this letter we have studied the DM scattering cross section in the simplest Higgs-portal DM model.
We have also discussed the limits required for direct detection in terms of the astrophysical observations on LSS. 
We observe that $(a)$ in compared with the future LHC with extremely large integrated luminosity $\mathcal{L}$ (at least of order $10^{3}$ $\text{fb}^{-1}$)
astrophysical observations on LSS provides a complementary way to detect the resonant mass region,
which is beyond the reach of Xenon 1T experiment;
$(b)$ the large mass region with $m_s$ above 3 TeV is beyond the reaches of all direct detections from the LSS, LHC and Xenon 1T. \\

~~~~~~~~~~~~~~~~~~~~~~~~~~~~~~~~~
$\mathbf{Acknowledgments}$\\
We would like to thank the referee for comments. 
This work is supported in part by Natural Science Foundation of China under Grant No.11247031 and 11405015.

\end{document}